\documentclass[12pt]{article}
\usepackage{epsf,aaspp4,flushrt}
\begin{document}
 
\def\cf{{\it cf.~}\ }
\def\deg{\ifmmode {^{\circ}}\else {$^\circ$}\fi}
\def\ergps{\ifmmode {\rm\,erg\,s^{-1}}\else ${\rm\,erg\,s^{-1}}$\fi}
\def\ergpscm2{\ifmmode {\rm\,erg\,s^{-1}\,cm^{-2}}\else
    ${\rm\,erg\,s^{-1}\,cm^{-2}}$\fi}
\def\ergpscm2A{\rm\,erg\,s^{-1}\,cm^{-2},{\AA}^{-1}}
\def\ergpsHz{\ifmmode {\rm\,erg\,s^{-1}\,{Hz}\,s^{-1}}\else 
    ${\rm\,erg\,s^{-1}\,{Hz}\,s^{-1}}$\fi}
\def\etal{{et al.~}}
\def\eg{{e.g.~}}
\def\Ho{\ifmmode {\rm\,H_0}\else ${\rm\,H_0}$\fi}
\def\ie{{\it i.e.~}\ }
\def\kms{\ifmmode {\rm\,km\,s^{-1}}\else
    ${\rm\,km\,s^{-1}}$\fi}
\def\kpc{{\rm\,kpc}}
\def\lsun{{\rm\,L_\odot}}
\def\lya{{\rm\,Ly$\alpha$~}}
\def\msun{{\rm\,M_\odot}}
\def\q0{\ifmmode {\rm\,q_0}\else ${\rm q_0}$\fi}
\def\spose#1{\hbox to 0pt{#1\hss}}
\def\simlt{\mathrel{\spose{\lower 3pt\hbox{$\mathchar"218$}}
     \raise 2.0pt\hbox{$\mathchar"13C$}}}
\def\simgt{\mathrel{\spose{\lower 3pt\hbox{$\mathchar"218$}}
     \raise 2.0pt\hbox{$\mathchar"13E$}}}
\def\ltsima{$\; \buildrel < \over \sim \;$}
\def\gtsima{$\; \buildrel > \over \sim \;$}

\def\refindent{\par\noindent\parskip=2pt\hangindent=3pc\hangafter=1 }

%

\def\refp#1#2#3#4{\refindent{#1,} {#2}, #3, #4}
\def\refb#1#2#3{\refindent{#1}{ {#2}, }{#3}}
\def\refx#1{\refindent{#1}}

\topmargin -0.4in
\oddsidemargin 0.0in
\evensidemargin 0.0in
\textheight 9.0in
\textwidth 6.5in

\title{Morphological Evolution in High Redshift Radio Galaxies and
the Formation of Giant Elliptical Galaxies\altaffilmark{1}}

\bigskip
\bigskip
\author{Wil J. M. van Breugel \& S. A. Stanford}
\affil{Institute of Geophysics \& Planetary Physics, Lawrence Livermore
National Laboratory, Livermore, CA 94550}
\affil{wil@igpp.llnl.gov, adam@igpp.llnl.gov}
\bigskip
\author{Hyron Spinrad, Daniel Stern, \& James R. Graham}
\affil{Astronomy Department, University of California, Berkeley,
CA 94720}
\affil{spinrad@bigz.berkeley.edu, dan@copacabana.berkeley.edu, jrg@graham.berkeley.edu}

\bigskip
\bigskip
\bigskip
\bigskip

\centerline{Submitted to The Astrophysical Journal on 11 August 1997}

\vskip 0.1in
\received{}
\accepted{}

\altaffiltext{1}{Based on observations obtained at 
the W.\ M.\ Keck Observatory, which is operated jointly by the 
University of California and the California Institute of Technology.}

\newpage

\begin{abstract}

We present deep near--infrared images of high redshift radio galaxies
(HzRGs) obtained with the Near Infrared Camera (NIRC) on the Keck I
telescope.  In most cases, the near--IR data sample rest wavelengths
free of contamination from strong emission lines and at $\lambda_{\rm
rest} > 4000$\AA, where older stellar populations, if present, might
dominate the observed flux.  At $z > 3$, the rest--frame optical
morphologies generally have faint, large--scale ($\sim$50 kpc) emission
surrounding multiple, $\sim 10$ \kpc~size components. The brightest of
these components are often aligned with the radio structures.  These
morphologies change dramatically at $2 < z < 3$, where the $K$ images
show single, compact structures without bright, radio--aligned
features.

The linear sizes ($\sim 10$ \kpc) and luminosities ($M(B_{\rm rest})
\sim -20$ to $-22$) of the {\it individual} components in the $z > 3$
HzRGs are similar to the {\it total} sizes and luminosities of normal,
radio--quiet, star forming galaxies at $z = 3 - 4$ (Steidel et al.\
1996; Lowenthal et al.\ 1997).  For objects where such data are
available, our observations show that the line--free, near--IR colors
of the $z > 3$ galaxies are very blue, consistent with models in which
recent star formation dominates the observed light. Direct,
spectroscopic evidence for massive star formation in one of the $z >3$
HzRGs exists (4C41.17, Dey \etal 1997$a$).  Our results suggest that
the $z > 3$ HzRGs evolve into much more massive systems than the
radio--quiet galaxies and that they are qualitatively consistent with
models in which massive galaxies form in hierarchical fashion through the
merging of smaller star--forming systems.

The presence of relatively luminous sub--components along the
radio axes of the $z > 3$ galaxies suggests a causal connection with
the AGN.  We compare the radio and near--IR sizes as a function of
redshift and suggest that this parameter may be a measure of the degree
to which the radio sources have induced star formation in the parent
objects.  We also discuss the Hubble diagram of radio galaxies, the
possibility of a radio power dependence in the $K - z$ relation, and
its implications for radio galaxy formation.

Finally, we present for the first time in published format basic radio
and optical information on 3C~257 ($z=2.474$), the highest redshift
galaxy in the 3C sample and among the most powerful radio sources known.

\end{abstract}


\newpage

\section {Introduction}

Radio sources have allowed the identification of luminous galaxies out
to extremely high redshifts.  Optical/near--IR campaigns during the
past few years by several groups have resulted in the discovery of
more than 120 radio galaxies at $z > 2$, including 17 with $z > 3$,
and 3 with $z > 4$ (see De Breuck \etal 1997 for a recent summary).
At lower redshifts, powerful radio sources are consistently identified
with giant elliptical and cD galaxies (\eg Condon 1989), suggesting
that at the highest redshifts we are observing these massive galaxies
in their early stages of formation.  While recently developed
techniques of finding very distant star--forming galaxies (\eg $U$ and
$B$ band dropouts, Steidel \etal 1996) are yielding substantial galaxy
populations at $z \simgt 3$, radio galaxy samples remain the best
means of finding, and studying, the most massive galaxies at the
highest redshifts.  Hiearchical galaxy formation scenarios suggest
that these massive galaxies are assembled from smaller structures at
relatively late cosmic epochs (\eg Baron and White 1987).
Observations of HzRGs may provide a unique opportunity to study the
beginning of this process at the highest redshifts.

The parent galaxies of luminous radio sources (log $P_{\rm 21 cm} > 31
{\rm\,erg\,s^{-1}\,cm^{-2}}$) at low redshift are consistenly
identified with massive elliptical galaxies (\eg Condon 1989).
Detailed studies of nearby ellipticals, using both radio--quiet and
radio--loud selected samples, furthermore suggest that their radio
properties are, in a statistical sense, governed by the luminosities of
their host galaxies (Ledlow \& Owen 1995, 1996, and references
therein).  Confirming much earlier results by Auriemma \etal (1977),
Ledlow \& Owen conclude that the probability of detecting radio
emission increases with optical luminosity, and that radio source
activity is a transient and probably recurrent phenomenon which may
occur in all elliptical galaxies at some time during their evolution.
The rather dramatic difference in radio morphology between low
luminosity (edge-darkened, FR I type; Fanaroff \& Riley 1974) and high
luminosity (edge-brightened, classical--double, FR II type) radio
sources may be due to evolution of their parent galaxies and/or
environment (Ledlow \& Owen 1996), or to qualitative differences
between their central engines and their accretion rates (Rees 1982;
Baum, Zirbel, \& O'dea 1995).  The FR I / FR II dichotomy is probably
not fundamentally related to the stellar content and mass of their
parent galaxies.  However, there are strong indications that the host
galaxies of powerful FR II sources are most often associated with
ellipticals that show some evidence for recent galaxy interactions.
The dynamical disruption may force large--scale gas flows into the
nucleus, stimulating it to greater activity (Heckman \etal 1986;
Zirbel 1996).

Powerful radio galaxies, like other active galaxies, are relatively
rare (\eg Osterbrock 1989).  The local space density of radio loud
ellipticals (log $P_{\rm 21 cm} > 31 {\rm\,erg\,s^{-1}\,cm^{-2}}$) is
$< 0.1$\% that of normal galaxies (log $P_{\rm 21 cm} < 28
{\rm\,erg\,s^{-1}\,cm^{-2}}$; Condon 1989).  The utility of radio
emission, much like line emission (with which it is usually strongly
coupled), lies therefore in providing a convenient beacon for finding
galaxies over a large range in redshift.  If we assume that the
relationship between radio luminosity and galaxy type holds true at
high redshift, then radio surveys provide a means to systematically
study giant elliptical galaxy evolution over large lookback times.
Such an assumption is reasonable, given the surprisingly well--behaved
$K-z$ relation for radio galaxies at $0 < z < 4$ (Lilly and Longair
1984; Eales \etal 1997), and the discovery of several radio
galaxies at moderate redshift having ages consistent with an early
formation epoch similar to that inferred for elliptical galaxies
(Dunlop \etal 1996; Spinrad \etal 1997; Dey \etal 1997$b$).
Throughout this paper we will assume that the identification of
powerful radio sources with massive elliptical galaxies holds true at
high redshift.

Near--IR observations of HzRGs are useful for studying galaxy
evolution for two important reasons. First, they allow us to observe
the galaxies at the rest--frame optical ($\lambda_{\rm rest} >
4000$\AA) where older stellar populations, if present, may dominate
the observed emission.  Second, the rest--frame UV morphologies of
powerful radio galaxies at $z \gtrsim 0.8$ are aligned with the radio
emission, evidently affected by the parent active galactic nucleus
(AGN; McCarthy \etal 1987; Chambers, Miley, \& van Breugel 1987).  In
the rest--frame optical, however, the alignment effect at $z < 1.8$ is
less pronounced (Rigler \etal 1992; McCarthy 1993$b$), though still
present in some systems (Eisenhardt \& Chokshi 1990; Dunlop \& Peacock
1993).  The relatively quiescent $K-z$ diagram for HzRGs could be
explained if the rest--frame optical light in most of the $z < 1.8$
systems is dominated by a passively evolving stellar population formed
at $z_f \gtrsim 5$ (\eg Lilly \& Longair 1984; Lilly 1989).  This
scenario implies that the rest--frame optical morphologies of most $z
> 1.8$ HzRGs may be dominated by light from stellar populations, not
by scattered light or radiation from other non--stellar processes
related to the AGN.

With the 10~m Keck telescopes it is now possible to obtain near--IR
images with unprecedented sensitivity (23 mag arcsec$^{-2}$ at $K$ in
one hour at 4$\sigma$), allowing investigations of HzRGs in the
rest--frame optical to z $\sim$ 4.  In this paper we present deep
near--IR imaging of 15 HzRGs with $1.8 < z < 4.4$ to investigate the
evolution of radio galaxies at the highest redshifts.  Because of the
strong correlation between radio power and emission--line luminosity
(McCarthy 1993$a$; Zirbel \& Baum 1995), the HzRG in our sample are
likely to have strong [\ion{O}{2}], [\ion{O}{3}], and H$\alpha$ lines.
We have attempted to select galaxy/filter combinations that avoid
strong line emission redshifted into the near--IR bands, to ensure
that the observed light will be dominated by continuum emission.  In
\S 2 we present the observations, followed by a description of the
results for individual systems in \S 3, and a discussion of
evolutionary trends apparent in the sample in \S 4.  In \S 5 we
discuss the $K - z$ diagram for HzRGs.  We also include an Appendix on
3C~257, providing basic optical and radio information on this powerful
and distant radio galaxy.

Throughout this paper we adopt H$_0 = 50$ km s$^{-1}$ Mpc$^{-1}$, q$_0
= 0.0$, and $\Lambda = 0$ for easy comparison to previously published
work on near--IR imaging of HzRGs ($i.e.$ Eales \etal 1997).  The
assumed cosmology implies a angular size scale of 13$-$14 kpc
arcsec$^{-1}$ for the redshift range $z = 1.8 - 4.4$ covered.  For
q$_0 = 0.1$ the corresponding angular size scale is 14\% -- 33\% less
at these redshifts.

\section {Observations and Data Reduction}

The observations described herein were made using NIRC (Matthews \etal
1994) at the Keck I telescope with mostly photometric conditions and
subarcsecond seeing ($\sim$0\farcs65 on average) in July 1994, January
1995, November 1996, and September 1997.  NIRC contains a
256$\times$256 InSb array with a scale of 0\farcs15 pix$^{-1}$.  The
target exposures were taken using a nonredundant dithering pattern,
with typical offsets between pointings of a few arcseconds.
Integration times for each pointing were typically 60--120s comprised
of 5--10 coadds, depending on the sky conditions and the wavelength of
the observations.  After bias subtraction, linearization, and
flatfielding, the data were sky--subtracted, registered and summed
using the DIMSUM\altaffilmark{1} near--IR data reduction package.  The
photometry was flux calibrated via short observations of UKIRT faint
standards, which, after a suitable transformation, yields magnitudes
on the CIT system (Casali \& Hawarden 1992).  A summary of the
observed objects, filters used, seeing, total integration times, and
magnitudes is given in Table~\ref{observations}.  Small and large
apertures were used to measure magnitudes; the former to obtain higher
signal--to--noise, and the latter for use in the $K-z$ diagram (see \S
5).  While the seeing varies from the $J$ to $K$ bands, the effect on
infrared aperture colors is small.  In Table~\ref{observations} we
note the rest wavelengths corresponding to the observed bands,
indicating when a strong emission line might contaminate the observed
flux.  The relevant radio parameters for our sources are presented in
Table~\ref{radiodata}.  While we include the IAU designations of the
objects in the latter table, in general we use the common names listed
in Table 1.

\altaffiltext{1}{DIMSUM is the Deep Infrared Mosaicing Software package,
developed by P.\ Eisenhardt, M.\ Dickinson, A.\ Stanford, and J.\ Ward,
which is available as a contributed package in IRAF.}

\section {Notes on Individual Galaxies}

Greyscale images of $12\arcsec \times 12\arcsec$ fields centered on
the observed radio galaxies are shown in Figure 1 (Plates 1 --- 4).
The objects are ordered in decreasing redshift, and rotated such that
the inner axes of the associated radio sources are aligned with the
horizontal to allow straightforward assessment of the degree of
radio/optical alignment (see Table~\ref{radiodata} for the radio
source position angles).  To the upper right of each figure we provide
a compass arrow indicating the cardinal directions.  Each panel
corresponds to an area of approximately $150\kpc \times 150\kpc$ at
these redshifts.  When an HzRG was observed in more than one band, we
typically only present either the most line--free image or the longest
wavelength image.  We wish to focus on the majority of the sample for
which the new data cover a spectral region above 4000\AA~in the
rest--frame at which stellar continuum emission, in particular from
older stars, when present, could dominate.  The panels show a trend of
decreasing overall complexity and size in the observed morphologies
with decreasing redshift.  The significance and possible
interpretation of this trend is considered in \S 4.1.  In this section
we discuss the HzRGs individually.

{\it 6C 0140+326 at z = 4.41} (Rawlings \etal 1996), is the highest
redshift radio galaxy known to date.  Our observations in September
1997 were made during photometric conditions and 0\farcs4 seeing.  The
image shows a faint, diffuse halo surrounding a bright,
multi--component and elongated feature which is comparable in length
($\sim$2\farcs5, $\sim 35\kpc$) and aligned with the radio structure.
Rawlings \etal suggest that the radio galaxy may be lensed by a
foreground galaxy (labelled G in our image; $z=0.927$) only
$\sim$1\farcs6 away.  Our $Ks$ image could contain [O
II]$\lambda$3727 at the blue edge of the filter, where it would be
passed at $\sim$20\% of the transmission peak.

{\it 8C 1435+63 at z=4.251} (Lacy \etal 1994), was observed in both
the $H$ and $K$ bands.  The image presented in Plate 1 is a sum of the
new 1920$s$ $K$ observation with the 3480$s$ $K$ observation reported
in Spinrad, Dey, \& Graham (1995, hereinafter SDG95). The system is
very faint and diffuse at $K$ with emission seen over a
$\sim$5$\arcsec$ (70 \kpc) long area.  The morphology in the $H$ band is
similar.  When measured over the central $3\arcsec$, the resulting
$H-K = 0.9$ is consistent with the prediction of a ``standard'' (1 Gyr
burst with solar metallicity and Salpeter IMF) passively--evolving
Bruzual \& Charlot (1997, hereinafter BC97) model for a formation
redshift $z_f = 7$ in our assumed cosmology.  $H-K$ is sensitive to
the age of a stellar population because it spans the 4000\AA~break at
$z=4.25$.  SDG95 argue that the $K$ light is likely due to
stars.  Problematic to such an interpretation, however, is the fact
that the near--IR morphology, like that of 6C~0140+326, is aligned
with the radio axis, implying an unlikely connection between a
moderately old stellar population (1.2 Gyr according to the BC97
model) and a short--lived radio source (typical ages $\sim 10^{7}$
yr).

{\it 4C 41.17 at z=3.800} (Chambers, Miley, and van Breugel 1990;
Chambers \etal 1996$ab$), with emission at $K_s$ spread nonuniformly
over a $\sim 3\arcsec \times 6\arcsec$ ($42 \times 84\kpc$) area, is
one of the largest HzRGs.  We observed 4C~41.17 in the $J$ band to
compliment the $K_s$ NIRC imaging reported in Graham \etal (1994).
The morphologies are very similar and allow us to study the $J-K_s$
colors of 4C~41.17.  The $J$ and K$_s$ bands are free of strong
emission lines, while $H$ and $K$ are contaminated by line emission
from [\ion{O}{2}] and H$\beta$$+$[\ion{O}{3}], respectively.  We find
that both the radio aligned component, 4C~41.17--North (N in Plate 1),
and its diffuse southern component, 4C~41.17--South (S in Plate 1; see
also van Breugel et al.\ 1997, 1998), are very blue with $J-K_s$ = 0.8
and 1.0, respectively.  Such blue colors at these redshifts are
consistent with the predictions of models for young star forming
systems (see also B2 0902+34, below).

{\it 4C 60.07 at z=3.790} (Chambers \etal 1996$ab$), was observed in
both $K$ and $K^\prime$.  The system has a complex $K^\prime$
morphology, with clumpy emission in the central area almost aligned
with the inner radio source (Chambers \etal 1996$b$), surrounded by a
diffuse, faint halo and a nearby companion, all spread out over a
$\sim$6\arcsec~(84~\kpc) long area.  The $K$-band image [not
shown] suffers from a small contribution from the [\ion{O}{3}] doublet
at the upper edge of the passband and has slightly more pronounced
filamentary features E and SW of the main body of the galaxy.

{\it MG 2144+1928 at z=3.594} (Maxfield \etal 1997), was observed in
the $H$ and $K$ bands to obtain color information, as well as the
$K^\prime$ band to examine the contamination of emission lines to the
other near--IR bands. It is one of the most elongated objects in our
sample with a maximum size of 8\arcsec~(111~\kpc).  The emission--line
free $K^\prime$ morphology is very diffuse, and aligned with the radio
axis.  Armus \etal (1997) present Keck/NIRC $K$ and narrow--band
2.3$\mu$m images of this galaxy, where the latter targets the
redshifted [\ion{O}{3}] emission line.  They find the galaxy to be
extended along the radio axis and estimate that 35\% of the total $K$
flux is contributed by the [\ion{O}{3}] emission line.

{\it 4C 1243+036 at z=3.581} (van Ojik \etal 1996), has unusually
diffuse IR emission aligned with the radio emission.  The galaxy was
observed in the $J$ and $K$ bands, as well as in a narrow--band ``CO''
filter covering 2.28--2.31$\micron$.  The narrow--band image was used
to correct for any contamination from the [\ion{O}{3}] doublet, which
redshifts to 2.294 $\micron$ at $z=3.581$.  The [\ion{O}{3}] flux was
found to contribute only 10\%~of the observed $K$ band flux in the
main two parts of the system over a $\sim$2\arcsec $\times$ 4\arcsec~
area, indicating the $K$ morphology is dominated by continuum
radiation.  Both the $K$ band and narrow--band 2.3 $\micron$ images are
presented in Plate 2.  The object below the center may not be part of
the system as it shows no [\ion{O}{3}] emission. After correcting for
the [\ion{O}{3}] contribution, the morphology of the $z=3.581$ system
in the observed $K$ band is similar to that seen in the line--free $J$
band image.  The line--corrected colors of the radio galaxy are $J-K
\approx 2.3$.  The diffuse and faint emission extending over
$\sim$5\arcsec~(70\kpc) to the NE of the main system is likely
primarily due to continuum, since little [\ion{O}{3}] emission is
detected there.

{\it B2 0902+34 at z=3.395} (Lilly 1988), occupies a special place in
the history of HzRG studies as the first identified using near--IR
techniques and the $K-z$ diagram (Lilly 1988).  We observed it in the
$J$ band because that is the only near--IR band free of a strong
emission--line.  As discussed in Eisenhardt \& Dickinson (1992) and
Eales \etal (1993), the [O III] doublet contributes a major fraction
of the observed emission in the $K$ band, which suggests that the
observed $H$ band would be strongly contaminated by [O II] emission.
The $J$ band continuum morphology is very diffuse and clumpy,
comparable to that of the {\it HST} WFPC2 F602W image of Eisenhardt
\etal (1997) when the latter is smoothed to a similar resolution.  The
$J$ emission encompasses a $\sim 2.5\arcsec \times 5\arcsec$ area ($34
\times 55\kpc$), comparable to 4C~41.17, 4C~60.07, and 4C~1243+036.
However, in B2~0902+34 there is no evidence for a strong
radio--aligned component.  This reinforces the suggestion by
Eisenhardt \& Dickinson (1993), on the basis of its line corrected
$R-K = 1.9$ and large diffuse structure, that B2~0902+34 is a young
forming galaxy in which most of the rest--frame optical light is
unaffected by the radio source.

{\it B3 0744+464 at z=2.926} (McCarthy 1991), has a regular and
compact structure in the $K_s$ image.  McCarthy classifies this object
as a broad--lined radio galaxy (BLRG) on the basis of broad \ion{C}{4}
and \ion{C}{3}] lines with narrow \lya and \ion{He}{2} lines.
Unfortunately, while the $J$ band imaging was obtained in the very
good seeing (0\farcs5), the $K_s$ band imaging suffers from by far the
worst seeing (1\farcs3) of our data.  The $J$ magnitude presented in
Table~\ref{observations} was measured after convolution with a
Gaussian to approximate the resolution of the $K_s$ band image.
Although a de Vaucouleurs profile can be fit to the original $J$ band
image, the profile appears nearly stellar (Figure~\ref{surfbright}$c$;
see below).  Given its very blue $J - K_s \sim 1.4$, BLRG
spectroscopic classification, and compact structure, B3~0744+464
appears to be dominated even in the near--IR by light from a strong
AGN.

{\it MRC 0943$-$242 at z=2.922} (McCarthy \etal 1996; also TX
0943-242, R\"ottgering \etal 1995), presents a rest--frame optical
structure with a centrally concentrated morphology.  As seen in
Figure~\ref{surfbright}$b$, the surface brightness profile is somewhat
rounded, which is consistent with the late stages of dynamical
relaxation after a merger (Navarro \etal 1995).  The maximum angular
extent is $\sim$3\arcsec~(41~\kpc) in the observed $K$.

{\it 4C 28.58 at z=2.905} (Chambers \etal 1996$ab$), also has a more
compact appearance compared to most of the higher redshift objects,
with a maximum IR diameter of 3\arcsec~(41~\kpc).  The {\it HST} WF/PC
image shown in Miley (1992) and Chambers \etal (1996$b$) is clumpy,
consisting of two bright rest--frame UV knots, which are separated by
1\arcsec~and lie along the radio axis.  There is no clear one--to--one
correspondence with the rest--frame optical structure seen in our
near--IR image, except that it is generally aligned along the same
axis.  A small companion is seen $\sim$3\farcs1 to the NW.  Comparison
with the optical images presented in Chambers \etal (1996$b$) suggests
that this companion is very red.  The $K$ band profile of 4C~28.58may
be intrinsically similar to those of the other $z < 3$ HzRGs if the IR
morphology of its central region is dominated by the same compact
double system resolved in optical images by $HST$ imaging (Miley 1992;
Chambers et al.\ 1996$b$).

{\it MG 1019+0534 at z=2.765} (Dey, Spinrad, \& Dickinson 1995,
hereinafter DSD95), appears to consist of two galaxies engaged in a
close encounter, indicated by an A and B in Plate 3 (following the
nomenclature of DSD95).  However, spectroscopy suggests component B to
be a foreground object at $z=0.66,$ while component A is identified as
the HzRG (DSD95).  The maximum extent at $K$ of the radio galaxy is
$\sim$2\arcsec~(27~\kpc).  The surface brightness profile of the HzRG
was determined after the foreground galaxy was removed by the
subtraction of an artificial galaxy constructed from fitted ellipses.

{\it TX 2202+128 at z=2.704} (R\"ottgering \etal 1997), has a smooth,
compact morphology in the observed $K$ band image with a maximum size
of $\sim$2\arcsec~or 27~\kpc.  

{\it MRC 2025$-$218 at z=2.630} (McCarthy \etal 1996), was observed in
the $K^\prime$ band, to avoid including the redshifted H$\alpha$
emission line, and also in the $J$ band.  The morphology is compact
and fairly symmetric, except for a faint plume approximately along the
radio source axis.  The maximum size is 4\farcs1 (55~\kpc).  Previous
$JHK$ imaging by McCarthy, Persson, \& West (1992) found the near--IR
colors somewhat red compared to a model with exponentially-declining
star formation of time constant 1.5 Gyr (Bruzual 1983), but their low
S/N $JHK$ measurements may well be contaminated by [\ion{O}{2}],
[\ion{O}{3}]/H$\beta$, and H$\alpha$ respectively.  Our new data give
a $J-K^\prime = 1.1$ in the same 4\arcsec\ aperture, which is more
consistent with the predicted color of a young population.

{\it 3C 257 at z=2.474} is the highest redshift radio galaxy in the 3C
catalog.  The redshift, determined from optical spectroscopy obtained
by Spinrad and collaborators, is published herein for the first time
(see Appendix).  The observed $K$ band contains the $H\alpha$ line,
while the $H$ band contains the [\ion{O}{3}] doublet and the $J$ band
contains the [\ion{O}{2}] doublet.  Nevertheless, the $K$ (rest--frame
$\sim R$) band light is fairly smooth, with a maximum size of
$\sim$4\arcsec~(53~\kpc), and a morphology resembling that of an
elliptical galaxy.  The $H$ band image, taken in non-photometric
conditions, is still useful in determining whether there is any strong
morphological wavelength dependence in the rest--frame optical (see \S
4.2 below).

{\it MG 2121+1829 at z=1.861} (Stern \etal 1997), lies at the lower
limit of the redshift range under consideration.  The observed $K$
band samples a relatively red part of the optical rest--frame compared
to the higher redshift radio galaxies.  The morphology seen in Plate 3
has a maximum size of 2.5\arcsec~(32~\kpc), but is not as centrally
concentrated as the other $z < 3$ radio galaxies.

\section {HzRG Morphologies and Colors}

The first systematic near--IR observations of radio galaxies showed
that there is a surprisingly good correlation between $K$ band
magnitudes and redshifts (Lilly \& Longair 1984; Lilly 1989).  This
was interpreted as evidence that the rest--frame optical light in
these objects is dominated by stellar populations evolving passively
since their formation epoch at $z_f > 5$.  Subsequent multi--band
visual and near--IR photometry of larger samples of HzRGs, when
compared to the theoretical spectral energy distributions from stellar
population models, have supported the galaxy formation paradigm of
Eggen, Lynden--Bell, \& Sandage (1963; see $e.g.$ McCarthy \etal
1992).  The most direct estimates of high--redshift galaxy ages have
been made for two $z \sim 1.5$ low--power radio galaxies (Dunlop \etal
1996, Spinrad \etal 1997, Dey \etal 1997$b$).  When compared with a
wide variety of models, spectroscopic measurements yield stellar ages
of $\sim 4$ Gyr and $z_f > 5$ in these objects, indicating that
stellar populations in some galaxies formed at a time prior to the
observation epoch of even the $3 < z < 4$ radio galaxies in our
sample.

Our new imaging allows us to probe closer to the possible epoch of
radio galaxy formation at the rest--frame optical wavelengths where
relatively old stellar populations might dominate the observed
emission.  The sample presented in \S 3 allows us to begin addressing
important questions about the formation of the first radio galaxies:
1) when do the first massive galaxies appear?  2) do these massive
galaxies evolve into giant ellipticals?  3) is there evidence for
morphological evolution over the redshift range of our sample in their
rest--frame optical light?  and 4) does the $K-z$ relationship
continue at the highest redshifts $i.e.$ are the galaxies already on a
passive evolution track at z $\sim$ 4?

\subsection {Morphological Evolution:  Evidence of Massive Forming Ellipticals?}

Our observations show that at the highest redshifts, $z > 3$, the
rest--frame visual morphologies exhibit structure on at least two
different scales. They often consist of several relatively bright,
$\sim 1 \arcsec$--size ($\sim$10 \kpc) components, which are
surrounded by very extended ($\sim$50 $-$ 100 \kpc) diffuse emission
(particularly 4C~41.17, 4C~60.07, 4C~1243+036, and B2~0902+34). The
individual component luminosities are $M(B_{\rm rest}) \sim -20$ to
$-22$, and the brightest are often aligned with the radio sources.
The total HzRG luminosities, including the low surface brightness
large scale emission, are $M(B_{\rm rest}) \sim -25$ to $-26$ (Table
3).  For comparison, present--epoch L$_\star$ galaxies and, perhaps
more appropriately, ultraluminous infrared starburst galaxies, have on
average $M(B_{\rm rest}) \sim -21.0$ (Binggeli, Sandage \& Tamman
1988; Sanders \& Mirabel 1996), while brightest cluster galaxies in
the local universe have $M(B_{\rm rest}) \sim -23.0$.  For objects
where such data are available, our observations show that the
integrated, line--free colors ($R-K$, $J-K$, or $H-K$) of the $z > 3$
galaxies are very blue, consistent with the BC97 predictions for a
stellar population formed at $z_f = 7$.

The linear sizes and luminosities of {\it individual} HzRG components
are similar to the {\it total} sizes and luminosities of normal,
radio--quiet, galaxies at $z = 3 - 4$ which have been discovered in
the Hubble Deep Field ($HDF$, Williams \etal 1996) using the $U$ or
$B$-band dropout technique (Steidel \etal 1996; Lowenthal \etal 1997;
Trager \etal 1997). Most of these galaxies show subclumping at the
$HST$ resolution ($\sim 1\kpc$ at these redshifts), and have deduced
star formation rates of $\sim 5 - 25$ M$_{\sun}$ yr$^{-1}$.  Recent
$HST$ images of a number of HzRGs have shown that the individual $\sim
10$ \kpc~size HzRG components also are resolved into smaller,
\kpc~scale structures (van Breugel \etal 1997, 1998; Pentericci \etal
1998$a,b$; Chambers \etal 1996$b$; Miley 1992).

It appears therefore that the parent galaxies of $z > 3$ radio sources
have total sizes and luminosities which are several times those of
normal, radio--quiet galaxies at similar high redshifts, but also that
they show compact structures on similar linear scales.  These results
suggest that the $z > 3$ HzRGs evolve into much more massive systems
than the radio--quiet galaxies and that they are qualitatively
consistent with models in which such galaxies form in hierarchical
fashion through the merging of smaller star--forming systems.  A
possible problem with this scenario, if the $z > 3$ HzRGs are indeed
the progenitors of massive 10$^{12}$ M$_{\sun}$ ellipticals, is that
in standard hierarchical cold dark matter models such large systems
are thought to form relatively late, $i.e.$ at much lower redshifts
($e.g.$ Kauffmann, White, \& Guiderdoni 1993).

The presence of relatively luminous $M(B_{\rm rest}) \sim -20$ to $-22$
sub--components, which are aligned with the radio sources in most of
the $z > 3$ galaxies, suggests a causal connection with the AGN.  The
most popular explanations for such an alignment effect, which is most
prominent at rest--frame UV wavelengths, include induced star
formation, scattered light from a hidden quasar, and nebular continuum
emission (\eg McCarthy 1993; Dickson \etal 1995).  Detailed Keck
spectropolarimetry and $HST$ WFPC2 images of the radio--aligned
component in 4C~41.17--N at $z = 3.800$ have shown that the rest-frame
UV continuum is dominated by jet--induced star formation (Dey
\etal 1997$a$; van Breugel \etal 1997), and that the contribution from
any scattered AGN light must be small.  This suggests that
jet--induced star formation may also be an important process in the
other $z >3$ radio galaxies, and that even at the rest--frame optical
wavelengths of our observations ($\lambda_{\rm rest} \sim 4000 -
5000$\AA) the emission in the radio--aligned features may be dominated
by hot, young stars.

At lower redshifts, $z < 3$, the rest--frame optical morphologies
become smoother, smaller, more centrally concentrated, and less
aligned with the radio structure.  To quantify the visual contrast of
their morphologies with those of the $z > 3$ HzRG, we have calculated
the ratio of the average flux at $r = 5~\kpc$ and $r = 30~\kpc$ for
each of the HzRGs.  The average value of these contrast ratios is 4.7
at $z > 3$ and 19.1 at $z < 3$.  We have also attempted to describe
the IR sizes by measuring the area of the emission regions associated
with the HzRGs as shown in Figure 1 (except for 4C 41.17 where we used
the $K_s$ image).  The areas are listed in Tables 3 and 4; the average
surface brightnesses ($k$-corrected to the rest $B$ band; see below)
within this area for the $z > 3$ objects are also given in Table 3.
The observations of the higher redshift objects do not go deep enough,
relative to the surface brightness limit of the $z < 3$ observations,
to make up for the expected $(1 + z)^4$ cosmological dimming.
However, the point illustrated by the sizes in Table 3 vs those in
Table 4 is that the $z > 3$ objects are generally much larger than
those at $z < 3$.  If our imaging of the $z > 3$ objects were 
deeper, the measured sizes of these HzRGs would only become larger
still.  The two measures described here confirm the visual impression,
generated by inspection of Figure 1, that HzRGs undergo dramatic
morphological evolution with redshift in the rest frame optical.

This result could be subject to several selection effects.  The images
shown in Figure 1 sample differing rest--frame spectral regions as a
function of redshift, $i.e.$ $\sim B$ band at $z > 3$ and $\sim V$
band at $z < 3$. In present epoch galaxies, there is little difference between
rest--frame $B$ and $V$ band morphologies of massive galaxies.  In the
$z > 2$ HzRGs one might expect enhanced star formation would yield
increased morphological differences.  To explore this question, we
have observed two of the $2 < z < 3$ HzRGs both in the $K$ band and a
shorter wavelength infrared passband ($H$ for 3C~257, $J$ for
MRC~2025-218).  The shorter wavelength passbands sample a similar rest
wavelength as do the $K$ band images of the $z > 3$ HzRGs.  In
Figure 1d we present the multi--wavelength near--IR images
of these two galaxies.  The general similarity of each of these two $z
\sim 2.5$ HzRGs at rest $\sim B$ and $\sim V$, indicates that the
change with redshift in the HzRG near--IR images is not simply due to
a morphological $k$--correction.

We also should consider if a redshift--radio power dependence biases
our analysis and the inferred morphological evolution.  Most of the
365 MHz radio luminosities are within a factor of 2.5 of the mean
radio luminosity, log $L_{365} = 36.56$, where $L_{365}$ is the flux
at 365 MHz measured in the units of erg s$^{-1}$ Hz$^{-1}$.  The
average 365 MHz radio luminosity of the $z > 3$ and $z < 3$ radio
galaxies differs only by a factor of 4 (see Table~\ref{radiodata}).
This relatively small dispersion coupled with the profound
morphological differences between $e.g.$ 6C~0140+326 and
3C~257, both having log $L_{365} \sim 36.8$ but spanning a redshift range
of 2.5 to 4.4, indicates that the range in radio power for the sample
has relatively mild impact on the strong morphological evolution
reported.

We have also briefly investigated the possibility that the near--IR
compactness of the $z < 3$ HzRGs may be enhanced by relatively large
contributions from AGN (\eg see McCarthy \etal 1992).  None of the $z
< 3$ objects stands out in the $K-z$ diagram with a $K$ band excess
(see below), though the scatter is quite large, limiting the utility
of this test.  Assuming that the radio core fraction (CF in
Table~\ref{radiodata}), defined as the ratio of the core to total
radio flux density, is a measure of AGN activity, one might expect
that objects with relatively strong radio cores would be more highly
nucleated in the near--IR images. This trend is generally not
observed.  One apparent exception is B3~0744+464, which has a BLRG
spectrum (McCarthy 1991) and is very blue in $J - K_s$.  Its nearly
unresolved $J$ band morphology may be dominated by an AGN.
Higher spatial resolution observations with NICMOS on {\it HST} of
this and other compact HzRGs will help to determine the actual
AGN contribution in the rest frame optical.

We conclude that the evolution from large, diffuse, multi--component,
and predominantly aligned morphologies at $z > 3$ to centrally
concentrated elliptical-like galaxies at $z < 3$ suggests that at the
highest redshifts the radio sources more strongly affect the
appearances of their parent galaxies, possibly due to induced star
formation in the dense proto--galactic media.  At $z < 3$, this effect
becomes less important and the overall galaxy formation process of
merging sub--components results in the rounder, more symmetric
structures, at least when seen at rest--frame optical wavelengths.
This provides qualitative support for hierarchical galaxy evolution
models in which massive galaxies form through merging of sub--galactic
stellar systems (\eg Baron \& White 1987).  According to dynamical
simulations, soon after merging, the morphology becomes centrally
concentrated and similar to that of an elliptical galaxy (e.g.\ Barnes
\& Hernquist 1994).

\subsection {Surface Brightness Profiles}

Surface brightness profiles of the galaxies were measured in the
near--IR images using the STSDAS ellipse task.  The profiles are
generated by fitting elliptical isophotes to the data, with the
center, ellipticity and position angle of each ellipse allowed to
vary.  The iterative method used in the fitting of each ellipse is
described more fully in Jedrzejewski (1987).  The $z > 3$ surface
brightness profiles are shown in Figure~\ref{surfbright}$a$, except
for 6C~0140+326 and 4C~1243+363, where nearby bright objects caused
too much confusion to obtain useful profiles.  As already suggested by
their morphologies in Figure 1, the surface brightness profiles of the
$z > 3$ radio galaxies are fairly flat.  No attempt was made to fit
the $z > 3$ profiles with the standard galaxy functional forms.  All
of the profiles are resolved, with significant light well beyond the
radius of the seeing disk.  The FWHM of a Gaussian fit to a star in
the NIRC images ranged from 0\farcs4 $-$ 0\farcs8, except
for the $K$ band image of B3~0744+464.

The $z < 3$ host galaxy profiles in (Figure~\ref{surfbright}$b$) have
steeper profiles than those of the $z > 3$ radio galaxies.  We
attempted to fit the $z < 3$ surface brightness profiles with a de
Vaucouleurs r$^{1/4}$ law and with an exponential law, the forms
commonly used to fit elliptical and spiral galaxy profiles,
respectively.  We demonstrate the fitting for our best resolved object
at $z < 3$, 3C 257 at $z = 2.474$, in Figure 3.  Within the limited
dynamical range of the data, both functional forms fit the observed
profiles---neither is preferred.  In the case of B3~0744+464, shown in
Figure~\ref{surfbright}$c$, the $J$ band profile is indistinguishable
from that of a star.

To transform our observed near--IR magnitudes to the rest--frame $B$
(for $z > 3$) and $V$ ($z < 3$) bands we used the BC97 population
synthesis models.  The $k$--corrections for the observed
large--aperture $K$ band magnitudes were calculated using a model with
all the stars formed in a 1~Gyr burst, with solar metallicity and a
Salpeter initial mass function, for $z_f$ = 7.  The form of the
$k$--correction is described by Humason, Mayall, \& Sandage (1956).
Uncertainties in formation redshift and assumed cosmology result in
uncertainty in the $k$--correction.  For example, a 1 Gyr change in
model age for an object at $z = 3$ results in a $^{+0.1}_{-0.2}$ mag
change.  This uncertainty, however, grows very quickly with increasing
redshift, particularly for higher H$_0$ and/or q$_0$ cosmologies in
which the time between the formation epoch and the observed epoch is
severely attenuated.  B2~0902+34 was only observed in the $J$ band, so
there is a relatively larger uncertainty in the calculated $M_B$ due
to the larger wavelength separation between the observed and rest
frame bands.  The resulting estimates of the total rest--frame optical
luminosities of the HzRGs, shown in Tables 3 and 4, are 3--5
magnitudes more luminous than present epoch L$_\star$ galaxies, which
have M$_\star (B) \sim -21.0$ and M$_\star (V) \sim -21.9$ (Binggeli,
Sandage, \& Tamman 1988).

\subsection {Linear Radio/IR Size Evolution}

In Figure~\ref{radiooptsize} we plot the log of the ratio of the radio
and near--IR lengths against redshift.  The IR lengths are defined to
be the long axis of the $K$ band emission for those objects showing
the alignment effect, and the diameter for the circularly symmetric
objects.  The radio lengths are measured as the distance between the
two hot spots on either side of the nucleus.  For $z > 3,$ most of the
HzRGs have comparable radio and IR sizes.  The exceptions are 4C~41.17
and 4C~60.07, but even in these objects a large fraction of their
radio emission (60\% and 40\% respectively at 1.5 GHz) is confined
within their parent, elongated galaxies (Chambers \etal 1996$b$).
Although the sample is small and the uncertainties large it appears
that there is a tendency for the radio sources to be comparable in
size to their aligned galaxy components at $z >3$, and systematically
larger $relative$ to the host galaxies in the IR at $z < 3$, the
exception being MG 1019+0534 which is confused by a nearby object of
similar brightness in the observed IR.

A possible interpretation is that at $z > 3$ the radio sources are
strongly interacting with very dense ambient gas, triggering star
formation along the radio axis while in the more evolved and less
gas--rich galaxies at $z < 3$, such jet-induced star formation is much
weaker.  A strong radio/optical size anti--correlation has been found
in $z \sim 1$ radio galaxies (Best, Longair, \& R\"ottgering 1996).
Since at such relatively low-$z$ the galaxies are more--or--less
coeval, this has been interpreted as stemming from jet--induced star
formation and the subsequent dimming of the young stellar population
as the radio sources become older and larger.  Alternative
explanations which might apply are related to scattered light from
hidden quasars, which appear to be common in $z \sim 1$ radio galaxies
(\eg Cimatti \etal 1996, 1997; Dey \etal 1996).

\section {The HzRG K--z Diagram}

Recent near--IR observations of flux--limited samples of HzRGs with a
range in radio luminosities have been interpreted in terms of a
possible radio power--$K$ dependence in the $K - z$ diagram, as well
as an increase in the $K$ magnitude dispersion with redshift (Eales
\etal 1997).  Investigations of dependencies between radio and
optical/near--IR parameters (\eg radio power, spectral index, spectral
curvature; accuracy of radio/optical/near--IR alignments; the
fractional contribution of the aligned UV flux to the flux at 5000\AA)
have resulted in a number of possible correlations, as well as
speculative interpretations (Lilly 1989; Dunlop \& Peacock 1993; Eales
\etal 1997).  General agreement, however, seems to exist that the
strength of the alignment effect depends on the power of the radio
sources and that the most luminous radio galaxies may be too extreme
in their rest--frame optical properties to allow an unbiased study of
their stellar populations in the $K$ band, while lower power HzRGs may
inhabit galaxies which are less affected by their resident AGN.  We
turn next to the Hubble diagram to further compare the low and high
power HzRGs.

A $K - z$ diagram is presented in Figure~\ref{kz}, combining our new
$K$ band photometry with 3C and 6C data from Eales \etal (1997).  We
have followed the same photometric procedures described in Eales \etal
to ensure that all magnitudes are on the same metric system.  There is
some uncertainty in the highest redshift point (6C~0140+326) because
of possible gravitational lensing (Rawlings \etal 1996).  As for the
rest of the objects in our sample, only 3 of the 14 HzRGs in our
sample with $K$ band data suffer from serious line emission
contamination which might affect their placement in the $K - z$
diagram.  It is interesting to note that for 3C~257, which most
diverges from the Hubble relation in our sample, Eales \& Rawlings
(1996) estimate a $\sim$35\% emission line contribution to the $K$
flux.  One could imagine that, as in the case of 3C~257, correcting
the $K$ magnitudes for strong line emission could lead to an even
tighter relation in the $K - z$ diagram.
 
Our new data fall on a reasonable extrapolation of the $K-z$ relation
as defined by the lower redshift data.  The 6C points appear to have a
slightly steeper slope in the $1 < z < 2$ range compared to the other
objects in Figure~\ref{kz}.  The radio luminosities of our NIRC sample
are comparable to those of the 6C sample reported in Eales \etal
(1997).  The main difference between the two samples is that nearly
all our objects are from samples which were selected to have radio
sources with steep radio spectra (``red radio colors'', with $\alpha_R
< -1$; $S_\nu \sim \nu^{\alpha_R}$).  Also, the 6C sample of Eales
\etal has a somewhat lower average redshift than the heterogeneous
sample defined by our NIRC data.

If the trend suggested by the slight divergence of the 6C and our NIRC
samples in the $K - z$ diagrams is real, it may be due to the
possibility that aligned structures in the $z > 3$ HzRGs may boost the
total star formation rates of their host galaxies during the earliest
stages.  In the case of 4C~41.17 (Dey et al.\ 1997), jet--induced star
formation may temporarily, during the lifetime of the radio source,
add as much as $\sim$30\% to the total star formation rate of the
entire galaxy system, and as much as $2 \times 10^9 \msun$ to the
total stellar population (van Breugel \etal 1998).  If radio source
activity is recurrent, as is likely since it is presumably related to
regular feeding of massive black holes through gas rich merger events,
then the cumulative effect of jet--induced star formation may add even
more to the total stellar population during the formative time of
these galaxies (the first $\sim$1 Gyr).  Correction of the $K$-band
values for the enhanced star formation by the radio sources might make
these 0.5---1 magnitudes fainter, driving them more in line with the
$K-z$ relation described by the lower redshift 6C objects.

Induced star formation will be significant only if large amounts of
cool gas in the form of dense clouds are encountered by the radio
source.  If this is not the case, for example because the
proto--galaxy is predominantly disk--shaped rather than spherical and
the radio-source axis is by chance directed at a large angle to this
disk, then star formation in this galaxy is little enhanced by the
radio source outflow (as may be true in B2 0902+34), compared to cases
where radio jets are propagating more within the galaxy disks
(possibly the case in 4C 41.17 and 4C 1243+036).  Furthermore, radio
power is at least indirectly related to the inter--cloud density of
its ambient medium because of the greater confinement pressure
provided (McCarthy, van Breugel, \& Kapahi 1991; Eales 1992).  Thus,
both the stellar luminosity of the galaxy and the power of the radio
source will depend on the morphology, density, and inhomogeneity of
the gaseous proto--galactic medium, as well as the age and duty cycle
of the radio source.  The result could be large enough variations in
the $K$ magnitudes of HzRGs to explain the dispersion, as well as a
possible radio power dependence, in the $K - z$ diagram.

\section {Conclusions}

Our near--IR observations of HzRGs show significant evolution in the
rest--frame optical ($\lambda_{\rm rest} \approx$ 4000---5000\AA)
morphologies of HzRGs.  At $z > 3$ the morphologies exhibit structure
on at least two different scales: relatively bright, compact
components with typical sizes of $\sim$10 \kpc~ surrounded by
large--scale ($\sim$50 $-$ 100 \kpc) faint and diffuse emission.  The
brightest components are often aligned with the radio sources, and
have individual luminosities $M(B_{\rm rest}) \sim -20$ to $-22$,
comparable to nearby L$_{\star}$ galaxies.  These morphologies change
considerably towards lower redshifts, and at $2 < z < 3$ the HzRGs
have smooth and much more compact structures, with decreased
radio--optical alignment.  The lower redshift objects have shapes
resembling elliptical galaxies.  Analysis of their surface
brightness shows that our spatial resolution and dynamic range are
insufficient to clearly discriminate between elliptical (r$^{1/4}$) or
disk (exponential) profiles.  Because of the absence of strongly
radio--aligned, rest--frame optical features in the $z < 3$ HzRGs,
spectroscopic age determinations at $\lambda_{\rm rest} > 4000$\AA~of
the stellar populations in these systems may provide useful
constraints on the formation epoch of the stellar populations in these
massive systems.  Such observations may become feasible with the
advent of near--IR spectrographs on 10--m class telescopes.

We have found some evidence that most of the $z > 3$ radio sources and
their host galaxies are comparable in size and are aligned, while the
$z < 3$ radio sources appear systematically larger (in a relative
sense) and are not aligned.  We interpret this possible lack of a
correlation in the radio and optical size evolution, as well as the
simultaneous change from predominantly aligned to non--aligned
structure, as evidence that the $z > 3$ radio sources are interacting
strongly with the dense gaseous media of their forming galaxies,
boosting the overall star formation rate.  In the more evolved and
less gas rich ellipticals at $z < 3$ the radio sources do not affect
their stellar populations significantly.  We suggest that the
temporary effects of radio source activity, as well as inhomogeneities
in the gaseous media of their forming galaxies, might explain the
dispersion and possible radio luminosity dependence in the $K - z$
diagram at the highest redshifts.

\bigskip
\acknowledgements

This paper was written in part while WvB was on sabbatical leave
during January 1997 -- April 1997 at the Anglo-Australian Observatory,
the Australian National Telescope Facility, and the Mount Stromlo and
Sidings Springs Observatories. He appreciates the support provided by
these institutes. He thanks the Kookaburra's for their wake--up calls,
and his Australian colleagues for their warm hospitality and
invigorating discussions, with special thanks to Drs.\ J.\
Bland--Hawthorn, G.\ Bicknell, M.\ Dopita, and R.\ Sutherland.  We
further are grateful for valuable discussions with P.\ Eisenhardt, and
thank S. Rawlings abundantly for advance information regarding
6C~0140+326 and 8C~1435+635 and his work on the $K - z$ diagram.  We
thank Mark Dickinson, Jim Liebert, and Pat McCarthy for providing the
optical identification and redshift of 3C~257, Arjun Dey and Mike Liu
for obtaining the $H$ band observations of 3C~257, and Mike Liu for
his help in removing bleed trails in the NIRC image of 6C~0140+326.
Finally, we thank the referee for many useful suggestions which have
improved the text. The work by WvB and SAS at IGPP/LLNL was performed
under the auspices of the US Department of Energy under contract
W--7405--ENG--48.  The work of HS and DS was supported by NSF grant
443833--21713.

\newpage

\appendix
\section {Appendix}

3C~257, at $z=2.474,$ is the highest redshift galaxy in the 3C catalog
and among the highest radio luminosity sources known (see
Table~\ref{radiodata}).  We publish herein for the first time a radio
map, finding chart, and redshift for this interesting source.

A 0\farcs4 resolution radio map at 4885 MHz (6.1 cm) is shown in
Figure~\ref{3c257radio}.  The source was also observed with
1\farcs30 resolution at 1465 MHz (20.5 cm) and 4885 MHz (6.1 cm)
which we use to calculate an accurate radio spectral index.  These
observations reveal no additional radio structure but show that the two
lobes are very asymmetric in radio brightness and spectral index (see
Table~\ref{radio257a}), and a exhibit a large rest--frame rotation
measure for the S$_f$ lobe (see Table~\ref{radio257b}).  The radio lobe
separation is approximately 12\arcsec.

In Figure~\ref{3c257opt} we present a finding chart for 3C~257.  The
optical identification of the radio galaxy was done Spinrad and
collaborators and lies approximately midway between the radio lobes.
Typical of HzRGs, the optical (rest--frame UV) host galaxy is aligned
with the radio axis.  Optical spectroscopy and redshift identification
was also obtained by Spinrad and collaborators using the Lick
Observatory 3m, MMT, KPNO 4m, and Palomar 5m telescopes over a period
of several years.  In Table~\ref{3c257spec} we present the observed
emission lines and implied redshift.  The \lya line is broad and
diffuse spatially, with a flux of $9\times10^{-16}$ erg cm$^{-2}$
s$^{-1}$ and a rest--frame equivalent width of $\sim$100 \AA.

\newpage

\begin{deluxetable}{llcccccll}
\small
\tablewidth{7.0in}
\tablenum{1}
\tablecaption{Keck/NIRC Observations of High Redshift Radio Galaxies}
\tablehead{
\colhead{Object}&\colhead{z}&\colhead{Filter}&\colhead{Mag}&
\colhead{Mag}&\colhead{Seeing}&\colhead{Exp.}&
\colhead{Rest$\lambda\lambda$}&\colhead{Notes}\\
\colhead{}&\colhead{}&\colhead{}&\colhead{2\arcsec}&
\colhead{8\arcsec}&\colhead{\arcsec}&\colhead{s}&\colhead{}&\colhead{}
}
\startdata
6C 0140+326	& 4.41 	& $K_s$ & 20.7    & 20.0  & 0.41 & 3840 &3677-4486 & includes [O II] \nl
8C~1435+635	& 4.251	& $K$ & 20.1	& 19.5	& 0.6\tablenotemark{c} & 5760\tablenotemark{a} & 3809-4622 & \nl
		& 	& $H$ & 21.0	&\nodata& 0.6\tablenotemark{c} & 3600 & 2840-3474 & very faint\nl
4C~41.17	& 3.800 & $K_s$& 20.7	& 19.2	& 0.87 & 2372 & 4146-4833 & \nl
        	&       & $J$ & 21.1	&\nodata& 0.69 & 3300 & 2302-2910 & \nl
4C~60.07	& 3.790 & $K$ & 19.9	&\nodata& 0.65 & 2000 & 4175-5067 & [O III] at edge\nl
		&       & $K'$ & 20.1	& 19.3	& 0.65 & 3660 & 4081-4785 & \nl
MG~2144+192	& 3.594	& $K$ & 19.9	&\nodata& 0.62 & 3840 & 4354-5283 & includes [O III]\nl
        	&	& $K'$ & 20.4	& 19.2	& 0.65 & 2760 & 4256-4989 & \nl
		&	& $H$ & 20.5	&\nodata& 0.74 & 3600 & 3246-3970 & includes [O II]\nl
4C~1243+036 	& 3.581 & $K$ & 19.8\tablenotemark{b}& 19.3\tablenotemark{b}& 0.81 & 3000 & 4366-5298 & includes [O III]\nl
        	&       & CO  & \nodata &\nodata& 0.60 & 3000 & 4985-5045 & [O III] \nl
		&	& $J$ & 22.1	&\nodata& 0.63 & 3000 & 2412-3050 & \nl
B2~0902+34	& 3.395 & $J$ & 22.7:	& 21.3: & 0.73 & 4900 & 2514-3179 & very diffuse \nl
B3~0744+464	& 2.926	& $K_s$ & 18.9  & 18.5	& 1.30 & 600  & 5069-5909 & \nl
		&	& $J$ & 20.4	&\nodata& 0.50 & 1250 & 2815-3558 &\nl
MRC~0943$-$242 	& 2.922 & $K$ & 19.4	& 19.2	& 0.87 & 1000 & 5128-6223 & TX also \nl
4C~28.58    	& 2.905 & $K$ & 19.6	& 18.7	& 0.73 & 1800 & 5122-6215 & \nl
MG~1019+0534	& 2.765 & $K$ & 19.9	& 19.1	& 0.6\tablenotemark{c} & 2000 & 5312-6446 & superposition \nl
TX~2202+128 	& 2.704 & $K$ & 18.6	& 18.4	& 0.83 & 1320 & 5400-6552 & \nl
MRC~2025$-$218	& 2.630 & $K'$ & 19.0	& 18.5	& 0.77 & 2520 &5386-6314 & \nl
		&       & $J$ & 19.9    &\nodata& 0.48 & 1800 &3044-3719 & \nl
3C~257      	& 2.474 & $K$ & 18.4	& 17.8	& 0.75 & 2000 & 5757-6986 & includes H$\alpha$ \nl
         	&       & $H$ &\nodata	&\nodata& 0.79 & 1350 &4292-5181 & includes [O III] \nl
MG~2121+1839	& 1.861 & $K$ & 19.8	& 18.7	& 0.63 & 1560 & 6991-8483 & \nl
\enddata
\tablenotetext{a}{Total exposure after coadding new data with a 3480$s$ observation by SDG95.}
\tablenotetext{b}{Corrected for [\ion{O}{3}] emission (see \S 3).}
\tablenotetext{c}{Estimated seeing; no stars in field.}
\label{observations}
\end{deluxetable}

\clearpage

\begin{deluxetable}{lllllllllllc}
\tablewidth{0pt}
\tablenum{2}
\tablecaption{Radio Parameters of High Redshift Radio Galaxies}
\tablehead{
\colhead{Cat}&\colhead{Name}&\colhead{Common}&\colhead{$z$}&\colhead{S$_{365}$}&
\colhead{S$_{1400}$}&\colhead{$-\alpha_R$}&\colhead{CF}&\colhead{Size}&\colhead{PA}&
\colhead{Ref}&\colhead{log L$_{365}$}\\
\colhead{}&\colhead{J2000}&\colhead{Name}&\colhead{}&\colhead{mJy}&\colhead{mJy}&
\colhead{}&\colhead{\%}&\colhead{\arcsec}&\colhead{\deg}&\colhead{O,R}&
\colhead{}}
\startdata
6C &0143+3253 & 6C~0140+326 &4.41  & 451 & 92&1.19 &0.0 &2.6 &102 &Ra,Ra &36.73  \nl
8C &1436+6319 & 8C~1435+635 &4.251 &2823 &498&1.29 &3.3 &4.3 &155 &La,La &37.54  \nl
4C &0650+4130 & 4C~41.17    &3.800 &1113 &266&1.07 &0.6 &10  & 78 &Ch,Ca &36.81  \nl
4C &0512+6030 & 4C~60.07    &3.790 &1120 &157&1.46 &2.1 &9.0 & 98 &Ch,Ca &37.08  \nl
MG &2144+1929 & MG2144+1929&3.594 &1730 &344&1.20 &0.0 &8.9 &177 &Ma,Ca &37.01  \nl
4C &1245+0323 & 4C~1243+036 &3.581 &1947 &375&1.23 &1.0 &7.4 &145 &Oj,Oj &37.07\nl
B2 &0905+3407 & B2~0902+34  &3.395 &1129 &335&0.90 &6.6 &6.5 &142 &Li,Ca &36.55  \nl
B3 &0747+4618 & B3~0744+464 &2.926 &1851 &503&0.97 &1.2 &1.9 & 92 &M1,Ca &36.60 \nl
MRC&0945$-$2428& MRC0943$-$242&2.922 &1296 &272&1.16 &0.0 &3.9 & 73 &M2,Ca &36.56 \nl
4C &2351+2910 & 4C~28.58    &2.905 &1704 &262&1.39 &2.1 &17  &144 &Ch,Ca &36.80  \nl
MG &1019+0534 & MG1019+0534&2.765 & 925 &395&0.63 &5.8 &1.3 &105 &De,De &36.02  \nl
TX &2205+1305 & TX2202+128 &2.704 & 839 &198&1.07 &0.9 &4.2 & 72 &Ro,Ca &36.20 \nl
MRC&2027$-$2140& MRC2025$-$218&2.630 &1280 &342&0.98 &0.7 &5.1 & 17 &M2,Ca &36.30 \nl
3C &1123+0530 & 3C~257      &2.474 &5648&1784&0.86 &0.0 &13  &124 &Sp,vB &36.83  \nl
MG &2121+1839 & MG2121+1839&1.861 &922 &242&0.99 &0.0 &6   &145 &St,St &35.68  \nl
\enddata
\tablecomments{
Columns:
{\bf S$_{365}$} = radio flux density at 365 MHz (University of Texas
survey; Douglas \etal 1996);
{\bf S$_{1400}$} = radio flux density at 1400 MHz (NVSS; Condon et
al.\ 1997)
{\bf $\alpha_R$} = radio spectral index between 365 MHz and 1.4 GHz, $S_\nu \sim \nu^{\alpha_R}$;
{\bf CF} = radio core / total fraction;
{\bf Size} = maximum radio source size;
{\bf PA} = radio source postion angle.
{\bf Ref} = references to optical (O) and Radio (R) data:  Ca
= Carilli \etal 1997; Ch = Chambers \etal 1996$a,b$; De = Dey \etal
1995; La = Lacy \etal 1994; Li = Lilly 1988; 
Maxfield \etal 1997; M1 = McCarthy 1991; M2 =
McCarthy \etal 1996; Oj = van Ojik \etal 1996; Ro = R\"ottgering \etal
1997; Ra = Rawlings \etal 1996; Sp = Spinrad, this paper; St = Stern
\etal 1997; vB = van Breugel, this paper;
{\bf L$_{365}$} = radio luminosity at 365 MHz in ergs/s/Hz.}
\label{radiodata}
\end{deluxetable}

\clearpage

\begin{deluxetable}{lccc}
\small
\tablenum{3}
\tablecaption{Optical Properties of $z > 3$ Radio Galaxies}
\tablehead{
\colhead{Object } & \colhead{A\tablenotemark{a}} &  
\colhead{$M(B_{\rm rest})$} & \colhead{$\mu_B$\tablenotemark{b}} \\
\colhead{} & \colhead{(arcsec$^2$)} & 
\colhead{(mag)} & \colhead{(mag arcsec$^{-2}$)}
}
\startdata
6C~0140+326  & ~4.2 & $-$25.7 & 26.3 \nl
8C~1435+635  & 10.2 & $-$26.1 & 27.4 \nl
4C~41.17     & 15.3 & $-$25.7 & 27.4 \nl
4C~60.07     & 12.1 & $-$25.6 & 27.4 \nl
MG~2144+1929 & 10.4 & $-$25.5 & 27.2 \nl
4C~1243+036  & ~9.6 & $-$25.4 & 28.1 \nl
B2~0902+34   & ~8.2 & $-$26.0: & 26.3: \nl
\enddata

\tablenotetext{a}{surface area of object in $K$ band image (see text)}
\tablenotetext{b}{average surface brightness (within A), converted to
the rest-frame $B$ band(see text)}
\label{optbigz}
\end{deluxetable}

\begin{deluxetable}{lcc}
\small
\tablenum{4}
\tablecaption{Optical Properties of $z < 3$ Radio Galaxies}
\tablehead{
\colhead{Object} & \colhead{A\tablenotemark{a}} & 
\colhead{$M(V_{\rm rest})$}  \\
\colhead{} & \colhead{(arcsec$^2$)} & \colhead{(mag)} }
\startdata
B3~0744+464	& ~3.4\tablenotemark{b} & $-$26.0 \nl
MRC~0943$-$242	& ~4.7 & $-$25.2 \nl
4C~28.58	& ~6.0 & $-$25.7 \nl
MG~1019+0534	& ~4.5 & $-$25.1 \nl
TX~2202+128	& ~2.5 & $-$25.7 \nl
MRC~2025$-$218	& ~7.3 & $-$25.5 \nl
3C~257		& 10.1 & $-$26.0 \nl
MG~2121+1839	& ~4.1 & $-$24.1 \nl 
\enddata
\tablenotetext{a}{surface area in $K$ band image (see text)}
\tablenotetext{b}{measured in $J$ band image because of very poor seeing in
$K$ image}

\label{optlittlez}
\end{deluxetable}

\clearpage

\begin{deluxetable}{llll}
\small
\tablewidth{6.5in}
\tablenum{5$a$}
\tablecaption{Radio Properties of 3C~257: Total Intensity}
\tablehead{
\colhead{Region} & \colhead{S$_{\rm 20.5 cm}$} &
\colhead{S$_{\rm 6.1 cm}$} & \colhead{$-\alpha^{\rm 6.1cm}_{\rm 20.5 cm}$} \\
\colhead{} & \colhead{(mJy)} & \colhead{(mJy)} & \colhead{}
}
\startdata
N$_p$ lobe  & 1473 & 436 & 1.01 \nl
S$_f$ lobe  &  140 &  22 & 1.54 \nl
Integrated & 1619 & 459 & 1.05 \nl
\enddata
\label{radio257a}
\end{deluxetable}

\begin{deluxetable}{llllll}
\small
\tablewidth{6.5in}
\tablenum{5$b$}
\tablecaption{Radio Properties of 3C~257: Polarization}
\tablehead{
\colhead{Region} & \colhead{$\phi^\circ_{\rm 20.5 cm}$} & 
\colhead{$\phi^\circ_{\rm 6.1 cm}$} & \colhead{\%$_{\rm 20.5 cm}$} & 
\colhead{\%$_{\rm 6.1 cm}$} & \colhead{$RM$\tablenotemark{a}} \\
\colhead{} & \colhead{} & \colhead{} & \colhead{} & \colhead{} & 
\colhead{{(rad m$^{-2}$)}}
} 
\startdata
N$_p$ lobe  & 24 $\pm$ 12 & $-$16 $\pm$ 45 & 6.5 $\pm$ 3.8 &  9.7 $\pm$ 4.2 & -- \nl
S$_f$ lobe  & 20 $\pm$  3 & $-$58 $\pm$  2 & 6.7 $\pm$ 1.9 & 12.6 $\pm$ 2.8 & 434 \nl 
\enddata
\tablenotetext{a}{$RM = (1 + z)^2 \times (\phi^\circ_{\rm 20.5 cm} - \phi^\circ_{\rm 6.1 cm})/(\lambda_{\rm 20.5cm}^2 - \lambda_{\rm 6.1cm}^2) 
\label{radio257b}$ rad m$^{-2}$.}
\end{deluxetable}

\begin{deluxetable}{ccr}
\small
\tablewidth{6.5in}
\tablenum{5$c$}
\tablecaption{Optical Properties of 3C~257}
\tablehead{
\colhead{Line ID} & \colhead{$\lambda_{\rm obs}$} & \colhead{$z$} \\
\colhead{} & \colhead{(\AA)} & \colhead{}
}
\startdata
\lya 1216		& 4228 & 2.478 \nl
\ion{C}{4} 1549 	& 5380 & 2.472 \nl
\ion{He}{2} 1640	& 5700 & 2.476 \nl
\ion{C}{3}] 1909	& 6625 & 2.470 \nl
			&      & mean $z$ = 2.474 \nl
\enddata
\label{3c257spec}
\end{deluxetable}

\clearpage


\begin{figure}
\figurenum{1a}
\caption{\scriptsize{Near--IR images of HzRGs, presented in order of
decreasing redshift.  Each panel is 12\arcsec\ square, oriented such
that the inner radio axis (see Table~\ref{radiodata}) is parallel to
the abscissa.  This allows a direct appraisal of the alignment of the
radio and near--IR structures.  Approximately half of the images were
resampled at a smaller pixel scale during the reduction phase, thus
yielding artificially smoother images.  Typically the longest
wavelength, line--free image is presented for each galaxy.  The East
and North directions on the sky are indicated next to the upper left
corners of the images, with North shown by the heavy arrow.  The
lowest contour in each panel corresponds to the surface brightness
indicated by the number (in mag arcsec$^{-2}$) below these compass arrows.
\label{plates}}}
\end{figure}

\begin{figure}
\figurenum{1b}
\caption{(continued.)}
\end{figure}

\begin{figure}
\figurenum{1c}
\caption{(continued.)}
\end{figure}

\begin{figure}
\figurenum{1d}
\caption{(top) A comparison of the observed $K$ band and $H$ band
images of 3C 257; (bottom) and of the observed $K'$ band and $J$ band
images of MRC 2025-218. }
\end{figure}

\begin{figure}
\vspace{-2cm}
\figurenum{2}
\plotone{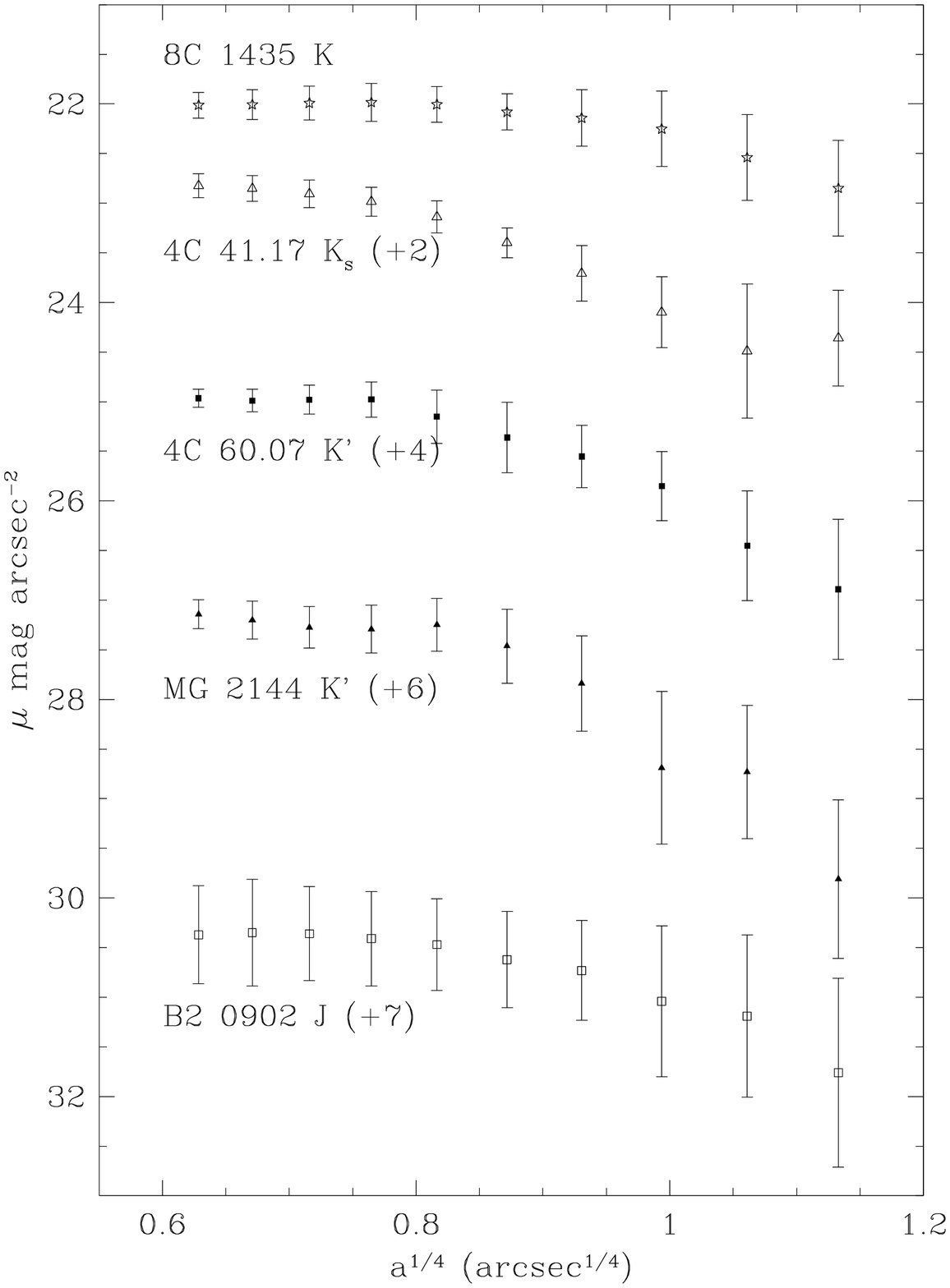}
\caption{Surface brightness profiles for the $z > 3$ objects are shown
in (a), for most of the $z < 3$ objects in (b), and for B3~0744+464
and 4C~28.58 in (c).  Seeing strongly affects the profiles generally
out to a radius of a$^{1/4} \sim$0.8 arcsec$^{1/4}.$ The profiles are
offset relative to the ordinate by the amount shown in parentheses in
each case.
\label{surfbright}}
\end{figure}

\begin{figure}
\figurenum{2}
\plotone{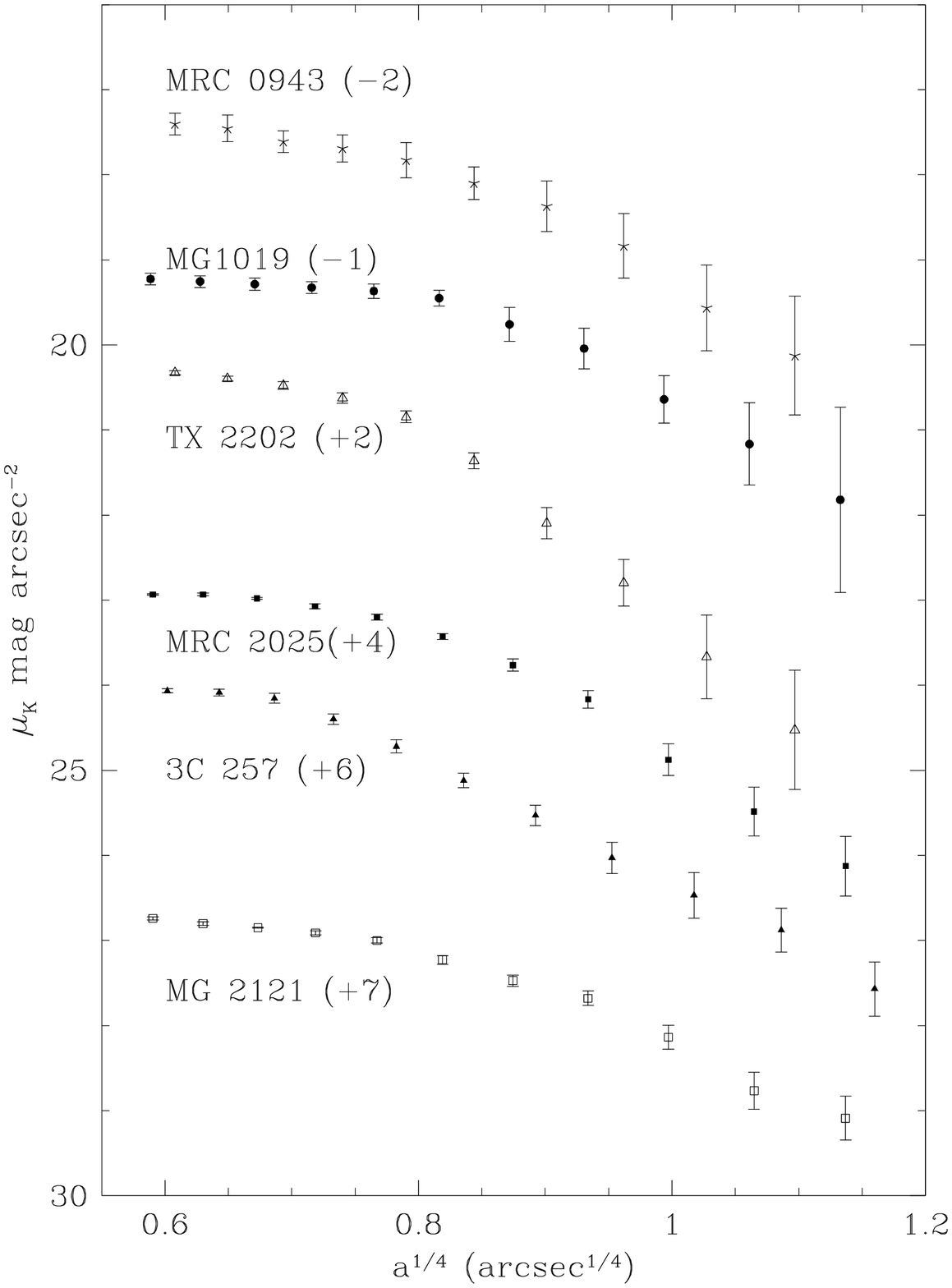}
\caption{(continued.)}
\end{figure}

\begin{figure}
\figurenum{2}
\plotone{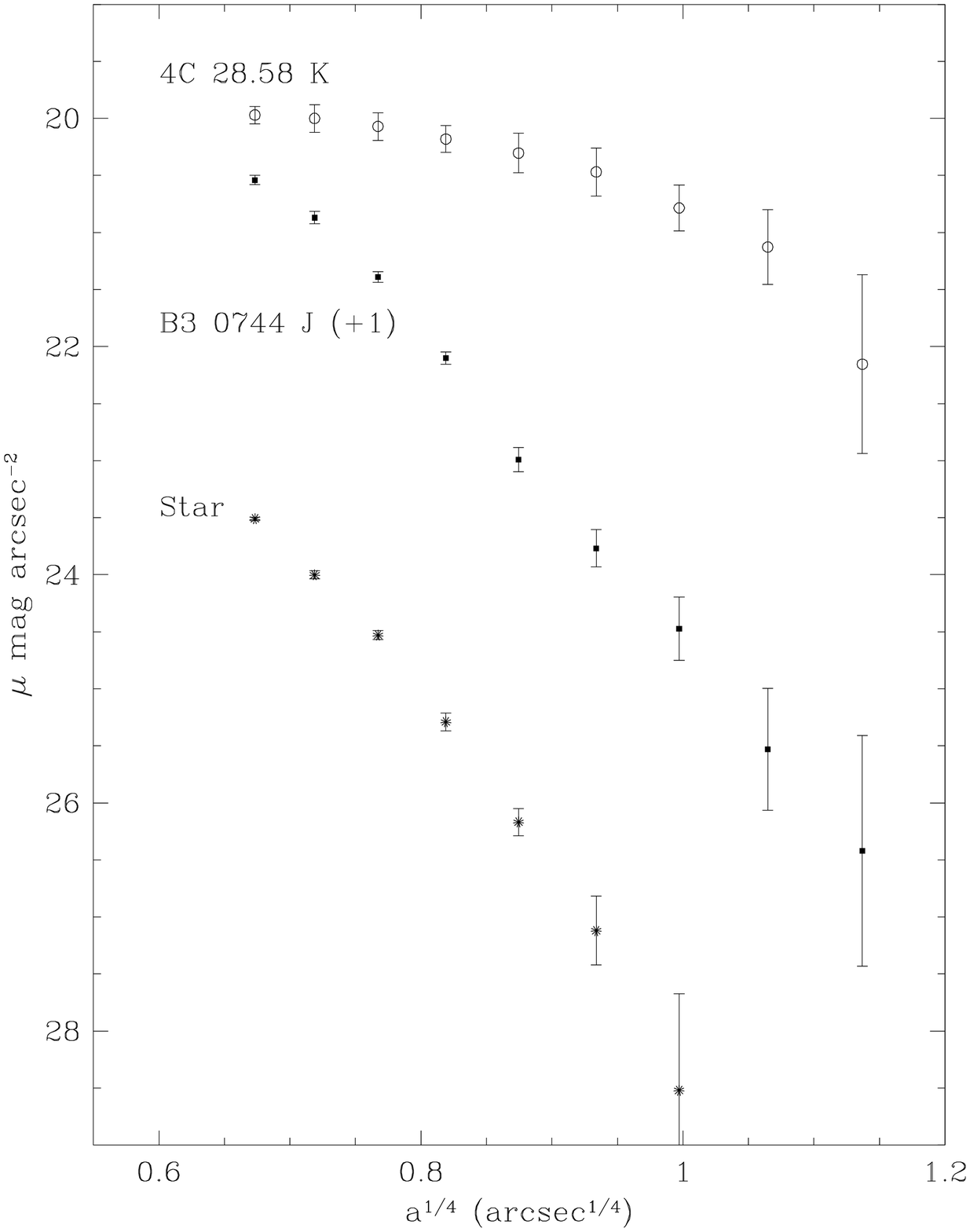}
\caption{(continued.)}
\end{figure}

\begin{figure}
\figurenum{3}
\plotone{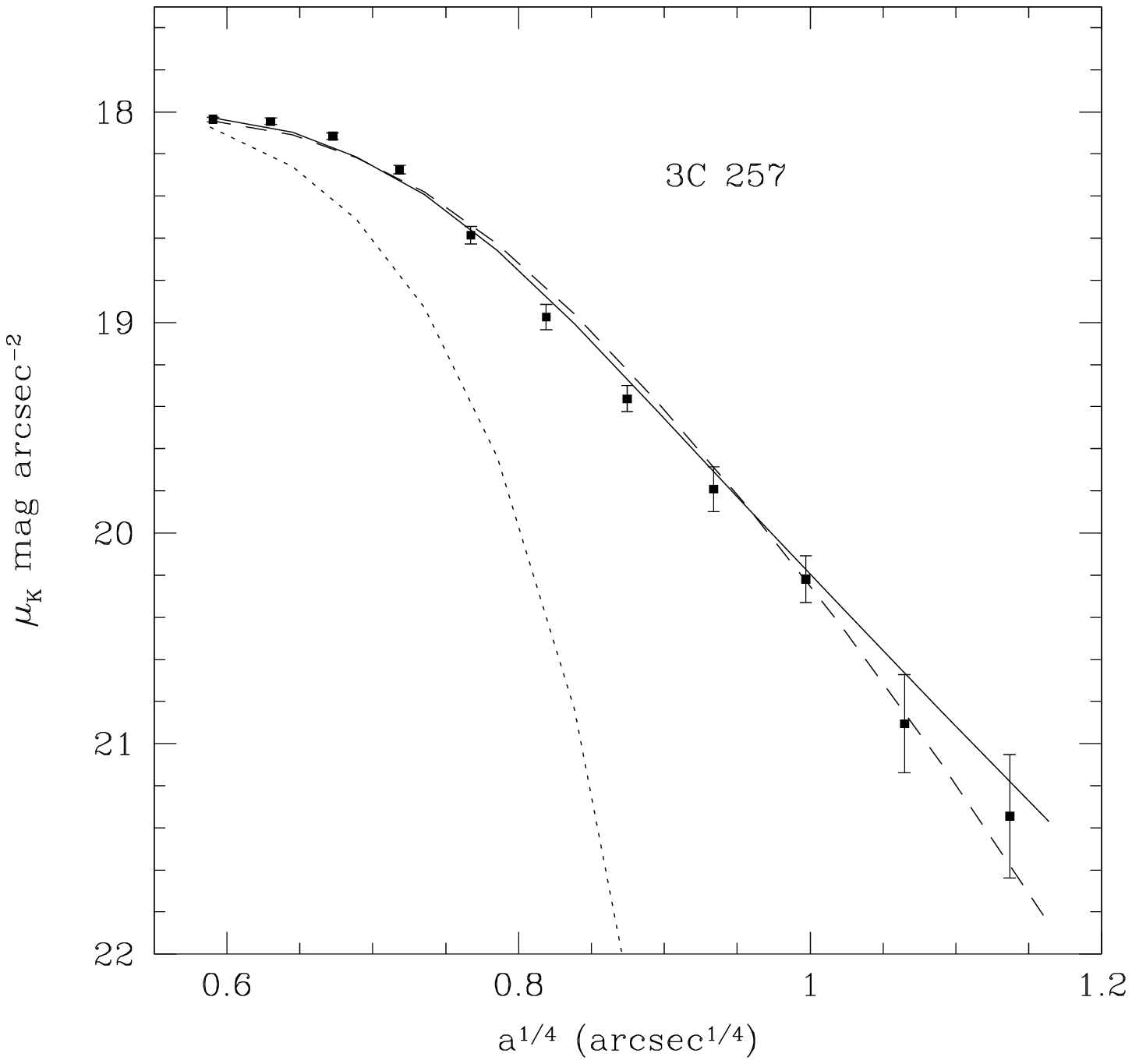}
\caption{The $K$ surface brightness profile of 3C~257, along with fits
by an exponential disk model (dashed line) and a de Vaucouluers model
(solid line).  The profile of a star in the $K$ image is shown by the
dotted line. \label{profits}}
\end{figure}

\begin{figure}
\figurenum{4}
\plotone{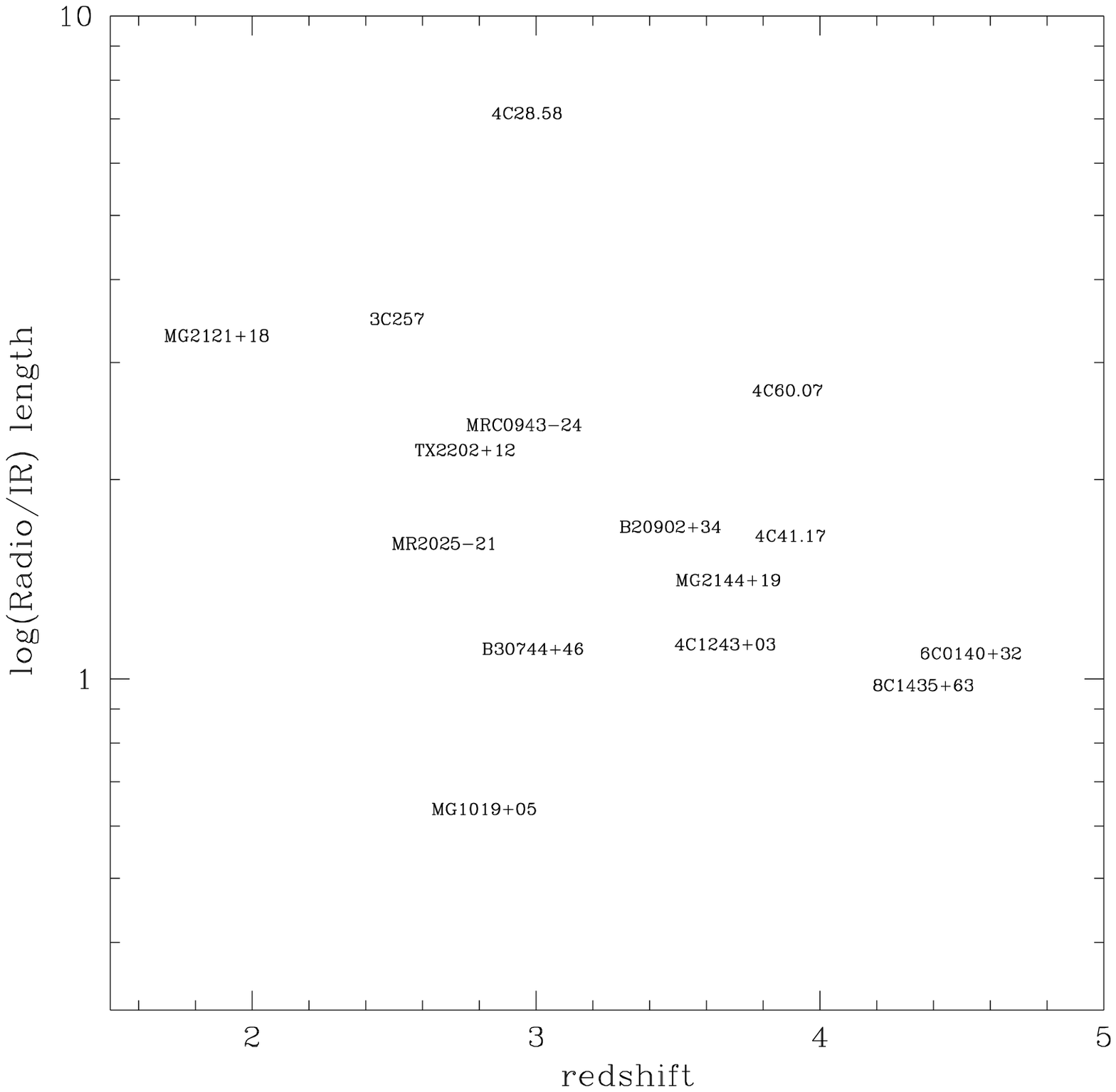}
\caption{The log of the ratio of the lengths in the
radio and in the near--IR vs redshift.  \label{radiooptsize}}
\end{figure}

\begin{figure}
\figurenum{5}
\plotone{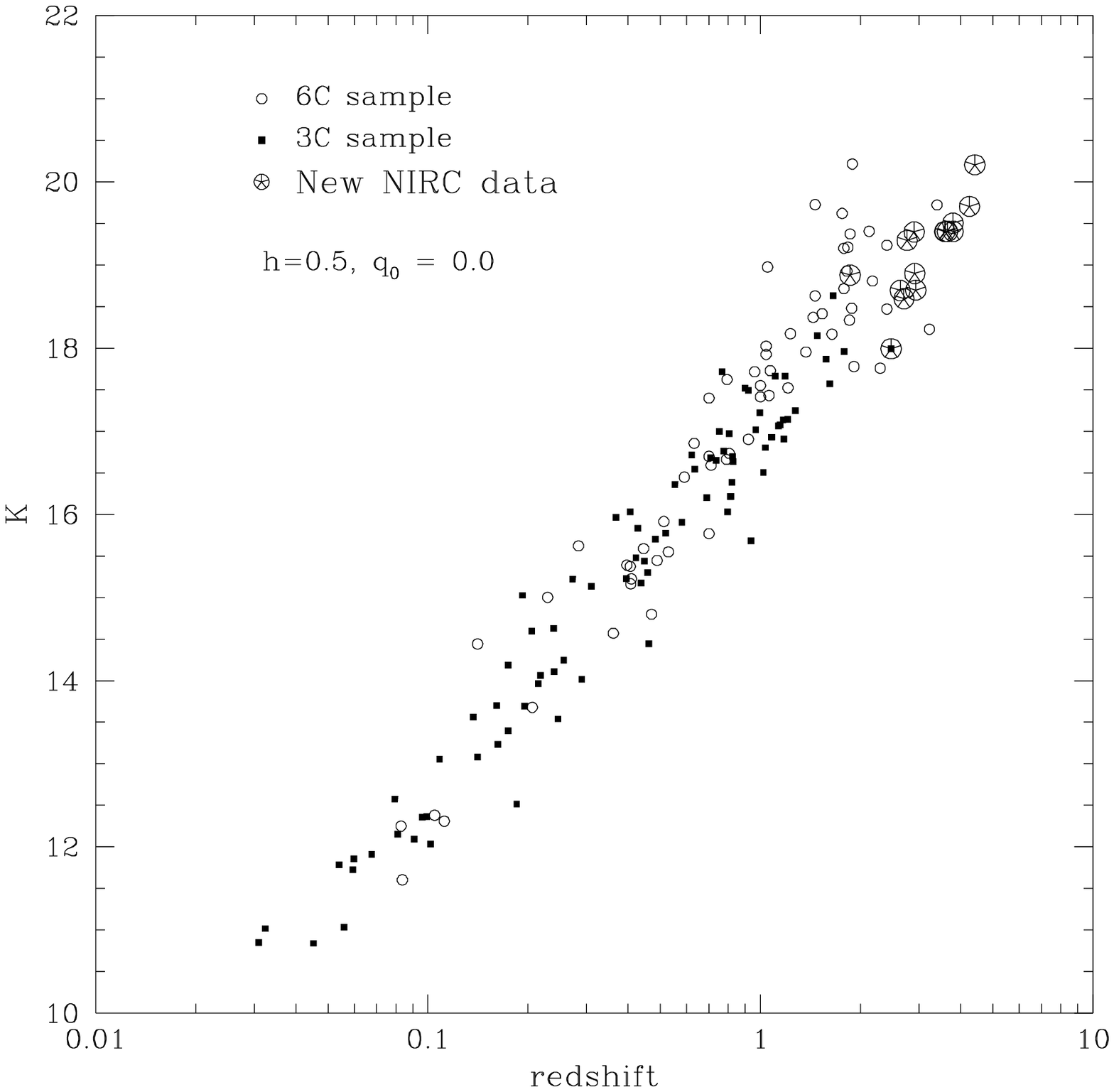}
\caption{Hubble $K-z$ diagram; the cartwheel points represent our new
sample, while all 
other photometry is from Eales et al.\ (1997). 
Magnitudes are aperture corrected to a 64 kpc metric diameter.
\label{kz}}
\end{figure}

\clearpage

\begin{figure}
\figurenum{6}
\plotone{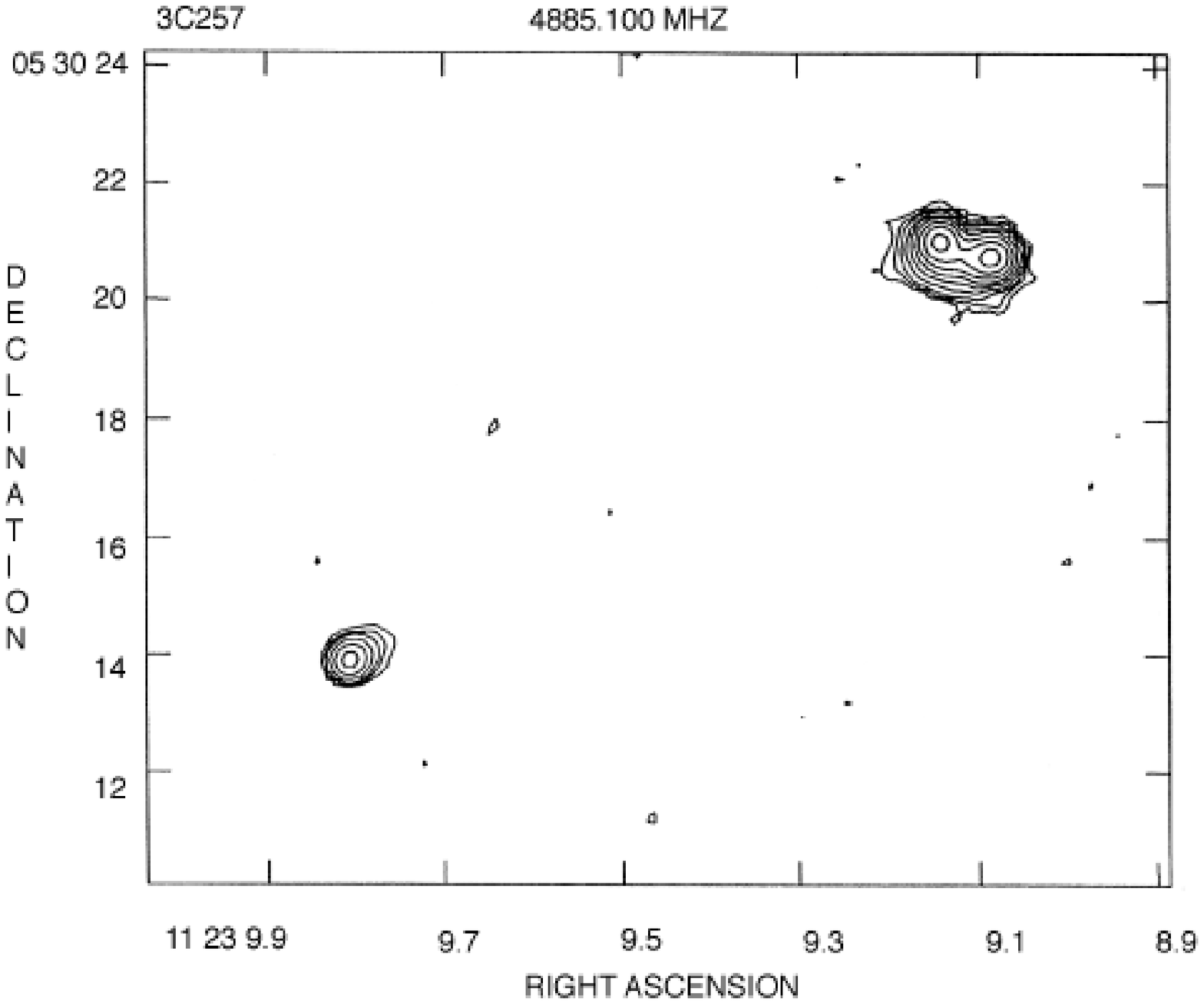}
\caption{Radio map of 3C~257 with 0.4\arcsec\ resolution at 4885 MHz
obtained at the VLA in the A--array. The source has no detected radio
core at a level of 0.9 mJy/beam (5$\sigma_{\rm rms}$) at this frequency.
Coordinates are in J2000. 
\label{3c257radio}}
\end{figure}

\clearpage

\begin{figure}
\vspace{-5cm}
\figurenum{7}
\plotone{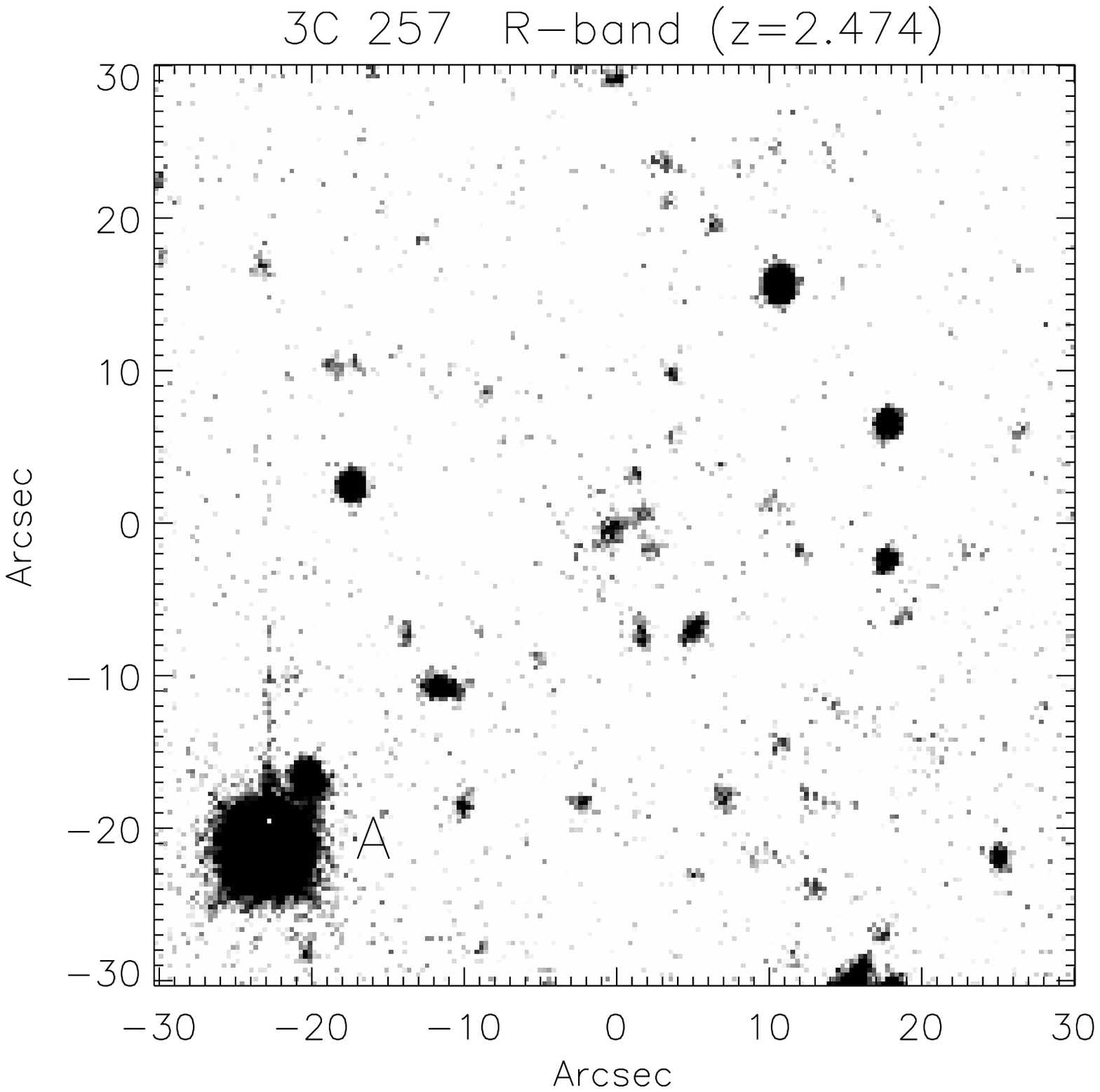}
\caption{$R$--band image of 3C~257 from the KPNO 4m telescope,
observed in March 1989 (Dickinson, priv. comm.).  North is to the top,
East is to the left. 3C257 is located at the center and the field is
60\arcsec\ on the side.  The optical emission is aligned with the
radio position angle ($PA = 124\arcdeg$), typical of HzRGs.  The
bright offset star, labelled `A' in the SE corner of the image, has
coordinates $\alpha_{J2000} = 11^h23^m10\fs9, \delta_{J2000} =
+5\arcdeg29\arcmin58\arcsec.$ The offset from star A to the radio
galaxy is $\Delta \alpha$ = -22\farcs7 (West), $\Delta \delta$ =
+20\farcs8 (North).
\label{3c257opt}}
\end{figure}

\end{document}